\documentstyle[aps]{revtex}

\begin{document}

\title{MAXWELL AND DIRAC THEORIES
AS AN ALREADY UNIFIED THEORY}

\author{Jayme Vaz, Jr.\thanks{vaz@ime.unicamp.br} and
Waldyr A. Rodrigues, Jr.\thanks{walrod@ime.unicamp.br}}

\address{Department of Applied Mathematics - IMECC\\
State University at Campinas (UNICAMP)\\
CP 6065, 13081-970, Campinas, S.P., Brazil}

\maketitle

\begin{abstract}
In this paper we formulate Maxwell and Dirac theories as an already
unified theory (in the sense of Misner and Wheeler).
We introduce Dirac spinors as ``Dirac square root'' of the Faraday
bivector, and use this in order to find a spinorial representation of
Maxwell equations.
Then we show that under certain circumstances this spinor
equation reduces to an equation formally identical to Dirac equation.
Finally we discuss certain conditions under which this equation can
be  really interpreted as Dirac equation, and some other possible
interpretations of this result.

\pacs{PACS Numbers: \/03.50De, 03.65.Bz, 03.65Pm}
\end{abstract}

\section{Introduction}

One very beautiful formal development concerning the structure of the
physical theories was given by the presentation of Misner and Wheeler
\cite{MW} of Maxwell and Einstein theories as an already unified
theory. Misner and Wheeler were able to describe the electromagnetic
field as a kind
of square root (the ``Maxwell square root'' in their terminology) of
the Ricci tensor, giving therefore a geometrical description of
electromagnetism by means of some algebraic conditions already
discovered by Rainich \cite{Rai}. This was the starting
pointing of Wheeler's program called geometrodynamics -- see, for
example \cite{Mielke}.

In this paper we shall consider a somewhat anologous construction for
the case of Maxwell and Dirac theories, that is, we try to formulate
electromagnetism and (relativistic) quantum mechanics as an already
unified theory. Using a terminology analogous to Misner and Wheeler,
we shall
introduce a Dirac spinor as a kind of square root -- the ``Dirac square
root'' -- of the electromagnetic field described by the Faraday
bivector. Moreover, we use this result in order to show that Maxwell
equations can be put in a form which is identical to Dirac equation.

First of all, before addresing our specifical problem, we shall introduce
the mathematical background used in this paper (sec.2). In particular,
we shall use in this paper the Clifford algebra formalism and the
concept of Dirac-Hestenes spinor. One can ask if there is a necessity
of doing this, if our calculations cannot be performed with usual
mathematical tools, etc. In this way we observe two things: firstly,
most of the original results of Misner and Wheeler were obtained due
to powerful
mathematical tools not widely known to physicists at that time
(1957), and, secondly, that Clifford algebras
seems to us to be a natural framework for physics, and in particular for
our problem; calculations are much more easy to be done using
Clifford algebras than with the traditional spinor and tensor
calculus -- an example of how powerful and natural they are is that
the use of the Clifford algebra of spacetime enable us to write
the Maxwell's eight scalar equations into a single equation\cite{He66}.
Moreover, our approach has a clear connection with quaternionic
quantum mechanics \cite{Adler,Leo}, and in this way we also clarify
some results of \cite{Leo}.
Anyway, our approach will be rather pedestrian, and one can easily find
rules for translating our results in terms of more usual concepts.

In sec.3 we introduce the Dirac-Hestenes spinor field as a  ``Dirac
square root'' of the Faraday bivector field -- this is possible once
we use a theorem of Rainich, Misner and Wheeler \cite{MW,Rai}.
Then we shall look for a spinorial representation of Maxwell
equations. There are in the literature several different spinorial
representations of Maxwell equations. These spinorial representations,
as a rule, use as components of the spinor field the components of
the electromagnetic field. The deficiency of this representations
is obvious: the action of the Lorentz group on the spinor field
does not give the correct transformation laws for the electric
and magnetic fields -- in fact, a spinor changes sign after a $2 \pi$
rotation, while the electromagnetic field components does not.
A detailed study of these kind of spinorial
representation of Maxwell equations can be found in \cite{Oli}.
The spinorial representation of Maxwell equations we shall use is
based on the concept of Dirac-Hestenes spinor, and does not suffer
the deficiency indicated above since the Faraday bivector is the
``Dirac square'' of the Dirac-Hestenes spinor.

Once we find a spinorial representation of Maxwell equations we
study it in details (sec.4), and we show that it can be reduced
to a form which is identical to Dirac equation; then we
study certain conditions under which this equation can be
really interpreted as Dirac equation. Other possible interpretations
are also discussed (sec.5), in particular those which may be relevant
to approaches to quantum mechanics based on the concept of quantum
potential.

\section{Mathematical Preliminaries: Clifford Algebras and
Dirac-Hestenes Spinors}

In this section we shall introduce the algebraic structures to be
used, and the concept of Dirac-Hestenes spinors in a pedestrian way.
We shall work with multivectors instead of multiforms, but the
translation is completely trivial for those used to work with the
latter.

\subsection{Exterior, Grassmann and Clifford Algebras}

Let $V$ be a vector space of dimension $n$ endowed with an interior
product $g : V \times V \rightarrow \mbox{${\sl  I\!\!R}$}$. Let
$\wedge$ be the exterior
(or wedge or Grassmann) product, that is, an associative, bilinear and
skew-symmetric product of vectors: $(a\wedge b)\wedge c = a\wedge(b\wedge c)$,
$(a + \alpha b)\wedge c = a\wedge c + \alpha b \wedge c $, $a \wedge b =
-b \wedge a$, ($\forall a,b,c \in V$). If $\{e_1 , \cdots , e_n\}$ is
a basis for $V$, then $\{e_1 \wedge e_2 , \cdots , e_1 \wedge e_n ;
e_2 \wedge e_3 , \cdots , e_{n-1}\wedge e_n \}$ is a basis for the
vector space $\bigwedge^2(V)$ whose elements are called bivectors
(2-vectors). In this way $\wedge : V \times V \rightarrow \bigwedge^2(V)$.
We can naturally extend the definition of the exterior product for vectors
and bivectors, giving trivectors (3-vectors), and so on. We denote by
$\bigwedge^k(V)$ ($0 \leq k \leq n$) the vector space of $k$-vectors,
which is of dimension $n \choose k$ (we adopt the convention $\bigwedge^0(V)
= \mbox{${\sl  I\!\!R}$}$ and $\bigwedge^1(V) = V$). We have $\wedge
: \bigwedge^k(V) \times
\bigwedge^l(V) \rightarrow \bigwedge^{k+l}(V)$, and if $A_k \in
\bigwedge^k(V)$ and  $A_l \in \bigwedge^l(V)$ then $A_k \wedge B_l =
(-1)^{kl} B_l \wedge A_k$. A $k$-vector is said to be simple if it is
the external product of $k$ 1-vectors. Consider the direct sum
${\displaystyle {\oplus_{k=0}^{n}}}\bigwedge^k(V) = \bigwedge(V)$.
An element of $\bigwedge(V)$ is called a multivector, and
$(\bigwedge(V),\wedge)$ is called exterior algebra. Let $\langle \hspace{1ex}
\rangle_k$ denote the projector $\langle \hspace{1ex}
\rangle_k : \bigwedge(V) \rightarrow \bigwedge^k(V)$. If $\langle A \rangle_k
= A_k$ the multivector $A$ is said to be homogeneous, and $k$ is its
grade.

We can extend $g$ to $\bigwedge(V)$. First, define the extension of
$g$ to $\bigwedge^k(V)$ as
\begin{equation}
\label{eq.0}
g_k(a_1 \wedge\cdots\wedge a_k, b_1
\wedge\cdots\wedge b_k) = {\rm det}\{g(a_i , b_j )\}
\end{equation}
for simple
$k$-vectors, and then extend to all $\bigwedge^k(V)$ by linearity
(by $g_0$ we mean ordinary multiplication of scalars). We denote
by $G$ the extension of $g$ to $\bigwedge(V)$ given by
$G = \sum_{k=0}^{n} g_k$, with $G(A_k , B_l) = 0$ when $k \neq l$.
($\bigwedge(V),\wedge,G$) is called Grassmann algebra. The structure
of Grassmann algebra is obviously richer than the one of exterior
algebra. In order to see this, let us first introduce two
involutions. The first, called reversion, and denoted by a tilde, is
defined by $\tilde{A}_k = (-1)^{k(k-1)/2}A_k$; the name reversion
is due to the fact that it reverses the order of the exterior
product of the vectors in a simple $k$-vector, i.e., $(a_1 \wedge \cdots
\wedge a_k)\tilde{} = a_k \wedge \cdots \wedge a_1 =
(-1)^{k(k-1)/2}(a_1 \wedge \cdots \wedge a_k)$. The second one, called
graded involution, and denoted by a hat, is defined by $\hat{A}_k =
(-1)^k A_k$. Now, let us define the contraction. The left contraction
${\mathbin{\rm\vphantom{X}\hskip 1.1pt\hskip -0.5pt\_\hskip -0.3em\_
\vrule width 0.5pt\hskip 1.7pt}}$ is defined by \cite{Lounesto}
\begin{equation}
\label{eq.1}
G(A {\mathbin{\rm\vphantom{X}\hskip 1.1pt\hskip -0.5pt\_\hskip -0.3em\_
\vrule width 0.5pt\hskip 1.7pt}} B,C) =
 G(B,\tilde{A}\wedge C) , \, \, \, \, \, \, \forall C \in \bigwedge(V) ,
\end{equation}
and the right contraction ${\mathbin{\rm\vphantom{X}\hskip
1.8pt\vrule width 0.5pt\hskip -0.7pt
\_\hskip -0.3em\hskip 0.8pt}} \, \,$ is defined by
\begin{equation}
\label{eq.2}
G(A {\mathbin{\rm\vphantom{X}\hskip 1.8pt\vrule width 0.5pt\hskip -0.7pt
\_\hskip -0.3em\hskip 0.8pt}}
 \, B, C) = G(A, C\wedge\tilde{B}) , \, \, \, \, \, \, \forall C \in
\bigwedge(V) .
\end{equation}
Left and right contractions are related by
\begin{equation}
\label{eq.3}
A_r {\mathbin{\rm\vphantom{X}\hskip 1.1pt\hskip -0.5pt\_\hskip -0.3em\_
\vrule width 0.5pt\hskip 1.7pt}} B_s =
 (-1)^{r(s-r)} B_s {\mathbin{\rm\vphantom{X}\hskip 1.8pt\vrule width
0.5pt\hskip -0.7pt
\_\hskip -0.3em\hskip 0.8pt}}
 \, A_r , \, \, \, \, \, \, (s > r) ,
\end{equation}
and satisfies the following properties:
\begin{equation}
\label{eq.4}
a {\mathbin{\rm\vphantom{X}\hskip 1.1pt\hskip -0.5pt\_\hskip -0.3em\_
\vrule width 0.5pt\hskip 1.7pt}} b = g(a,b) ,
\end{equation}
\begin{equation}
\label{eq.5}
a {\mathbin{\rm\vphantom{X}\hskip 1.1pt\hskip -0.5pt\_\hskip -0.3em\_
\vrule width 0.5pt\hskip 1.7pt}}
 (B\wedge C) =
(a {\mathbin{\rm\vphantom{X}\hskip 1.1pt\hskip -0.5pt\_\hskip -0.3em\_
\vrule width 0.5pt\hskip 1.7pt}} B)\wedge C
 + \hat{B} \wedge (a {\mathbin{\rm\vphantom{X}\hskip 1.1pt\hskip
-0.5pt\_\hskip -0.3em\_\vrule width 0.5pt\hskip 1.7pt}} C) ,
\end{equation}
\begin{equation}
\label{eq.6}
a {\mathbin{\rm\vphantom{X}\hskip 1.1pt\hskip -0.5pt\_\hskip
-0.3em\_\vrule width 0.5pt\hskip 1.7pt}}  (b
{\mathbin{\rm\vphantom{X}\hskip 1.1pt\hskip -0.5pt\_\hskip
-0.3em\_\vrule width 0.5pt\hskip 1.7pt}} C) = (a \wedge b)
{\mathbin{\rm\vphantom{X}\hskip 1.1pt\hskip -0.5pt\_\hskip
-0.3em\_\vrule width 0.5pt\hskip 1.7pt}} C  ,
\end{equation}
where $a,b \in V$, $B, C \in \bigwedge(V)$. In what follows, whenever
there is no danger of confusion, we denote the contraction by
a dot : $a \cdot B = a {\mathbin{\rm\vphantom{X}\hskip 1.1pt\hskip
-0.5pt\_\hskip -0.3em\_\vrule width 0.5pt\hskip 1.7pt}} B$ (the dot
must not be confused with
internal product in spite that for vectors we have eq.(\ref{eq.4})).

Let us introduce the Clifford (or geometrical) product $\vee$; given
$a \in V$ and $B \in \bigwedge(V)$ we define $a \vee B$ by
\begin{equation}
\label{eq.7}
a \vee B = a  {\mathbin{\rm\vphantom{X}\hskip 1.1pt\hskip
-0.5pt\_\hskip -0.3em\_\vrule width 0.5pt\hskip 1.7pt}} B +
a \wedge B  ,
\end{equation}
and extend this definition to all $\bigwedge(V)$ by associativity.
($\bigwedge(V),\wedge,G,\vee$) is called a Clifford algebra (denoted
by ${\mbox{${\it Cl}$}}(V,g)$). In order to simplify the notation, we denote
the Clifford product simply by justaposition, i.e.:
\begin{equation}
\label{eq.8}
aB = a \cdot B + a \wedge B  .
\end{equation}
Let $\{e_1 , \cdots , e_n\}$ be an orthonormal basis for $V$. In this
case we have
\begin{equation}
\label{eq.9}
e_{i}^{2} = e_i e_i = e_i \cdot e_i + e_i \wedge e_i = e_i \cdot e_i =
g(e_i , e_i) ,
\end{equation}
\begin{equation}
\label{eq.10}
e_{ij} = e_i e_j = e_i \cdot e_j + e_i \wedge e_j = e_i \wedge e_j
\, \, \, \, \, \, (i \neq j) .
\end{equation}
Since exterior, Grassmann and Clifford algebras are isomorphic as vector
spaces, a general element of ${\mbox{${\it Cl}$}}(V,g)$ is of the form
\begin{equation}
\label{eq.11}
A = a_0 + a^i e_i + a^{ij} e_{ij} + \cdots + a^{1\cdots n} e_{1\cdots n} ,
\end{equation}
where, in particular, $e_{1\cdots n} = e_1 \cdots e_n = e_1 \wedge \cdots
\wedge e_n = \tau$ is the volume element of $V$. We observe that
Clifford algebra is $Z_2$-graded algebra. In fact, let ${\mbox{${\it Cl}$}}_+$
(${\mbox{${\it Cl}$}}_{-}$) denote the set of elements of
${\mbox{${\it Cl}$}}$ with
even (odd) grade; we have ${\mbox{${\it Cl}$}}_{+}{\mbox{${\it
Cl}$}}_{+} \subset {\mbox{${\it Cl}$}}_{+}$,
${\mbox{${\it Cl}$}}_{+}{\mbox{${\it Cl}$}}_{-} \subset {\mbox{${\it
Cl}$}}_{-}$, ${\mbox{${\it Cl}$}}_{-}
{\mbox{${\it Cl}$}}_{+} \subset {\mbox{${\it Cl}$}}_{-}$,
${\mbox{${\it Cl}$}}_{-}{\mbox{${\it Cl}$}}_{-}
\subset {\mbox{${\it Cl}$}}_{+}$. This fact shows that ${\mbox{${\it
Cl}$}}_{+}$ is a
sub-algebra of ${\mbox{${\it Cl}$}}$, called the even sub-algebra.

The great advantage of Clifford algebra over Grassmann algebra follows
from the fact that the Clifford product contains more information
than Grassmann product. Moreover, under the Clifford product it is
possible to ``divide'' multivectors. In fact, define the
norm $\mid \! \! A \!\mid$ of a multivector $A$ as
\begin{equation}
\label{eq.12}
\mid \! A \! \mid^2 = \langle \tilde{A}A \rangle_0 .
\end{equation}
If $\tilde{A} A = \mid \! \! A \! \mid^2 \neq 0$ we define $A^{-1} = \tilde{A}
\mid \! \! A \! \mid^{-2}$; in fact $A^{-1}A = A A^{-1} = \mid \! A
\! \mid^{-2}
\tilde{A} A = 1$. Another advantage is that Clifford algebras
are isomorphic to matrix algebras (or direct sums)
over $\mbox{${\sl  I\!\!R}$}$,
$\mbox{${\sl  I\!\!\!\!C}$}$ or $\mbox{${\sl  I\!\!H}$}$
(we shall see examples below) \cite{Fig}.

Finally, we say that an algebraic element $e$ is an idempotent if
$e^2 = e$; it is called primitive if it cannot be written as the
sum of two mutually annihilating idempotents (that is: $e \neq e^{\prime}
+ e^{\prime\prime}$ with $(e^{\prime})^2 = e^{\prime}$, $(e^{\prime
\prime})^2 = e^{\prime\prime}$, $e^{\prime}e^{\prime\prime} =
e^{\prime\prime}e^{\prime} = 0$). The sub-algebra $I_E$ of an algebra $A$
is called a left ideal if given $i \in I_E$ we have $xi \in I_E$, $\forall
x \in A$ (similarly for right ideals). An ideal is said to be minimal
if it contains only trivial sub-ideals. Now, one can prove that
the minimal left ideals of a Clifford algebra ${\mbox{${\it
Cl}$}}(V,g)$ are of the
form ${\mbox{${\it Cl}$}}(V,g) e$, where $e$ is a primitive
idempotent \cite{Porteous}. This is
an important result to be used later.

\subsection{Spacetime Algebra}

Let $V = \mbox{${\sl  I\!\!R}$}^{1,3}$ be Minkowski vector space, and
choose a basis
$\{\Gamma_{\mu}\}$ ($\mu = 0,1,2,3$) such that $g(\Gamma_{\mu},\Gamma_{\nu})
= \eta_{\mu\nu} = {\rm diag}(1,-1,-1,-1)$. The spacetime algebra
(STA)\cite{He66}
is the Clifford algebra of $(\mbox{${\sl  I\!\!R}$}^{1,3},\eta)$,
denoted by $\mbox{${\sl  I\!\!R}$}_{1,3}$.
Observe that $\Gamma_{0}^{2} = -\Gamma_{1}^{2} = -\Gamma_{2}^{2} =
-\Gamma_{3}^{2} = 1$. A general element of $\mbox{${\sl
I\!\!R}$}_{1,3}$ is of the
form
\begin{equation}
\label{eq.13}
A = a + a^{\mu}\Gamma_{\mu} + \frac{a^{\mu\nu}}{2!}\Gamma_{\mu\nu} +
\frac{a^{\mu\nu\sigma}}{3!}\Gamma_{\mu\nu\sigma} + a^{0123}\Gamma_5  ,
\end{equation}
where $\Gamma_{\mu\nu} = \Gamma_{\mu}\wedge\Gamma_{\nu}$ ($\mu \neq \nu$)
and $\Gamma_5 = \Gamma_0 \Gamma_1 \Gamma_2 \Gamma_3 =
\Gamma_0 \wedge \Gamma_1 \wedge \Gamma_2 \wedge \Gamma_3 $ is the
volume element. Note that $\Gamma_{5}^{2} = -1$ and $\Gamma_{5} \Gamma_{\mu}
= -\Gamma_{\mu}\Gamma_{5}$.

The STA is particularly useful in formulating the Lorentz rotations.
Let $\mbox{${\sl  I\!\!R}$}_{1,3}^{+}$ denote the even sub-algebra of
$\mbox{${\sl  I\!\!R}$}_{1,3}$,
and let ${\cal N}$ be the norm map, i.e., ${\cal N}(L) = \mid \! \! L
\! \! \mid^2 = \langle \tilde{L}L \rangle_0$. The double covering of the
restricted Lorentz group ${\rm SO}_{+}(1,3)$ is ${\rm Spin}_{+}(1,3)$
defined as
\begin{equation}
\label{eq.14}
{\rm Spin}_{+}(1,3) = \{ R \in \mbox{${\sl  I\!\!R}$}_{1,3}^{+} \mid
{\cal N}(R) = 1 \} .
\end{equation}
An arbitrary Lorentz rotation is therefore given by $a \mapsto
RaR^{-1} = R a \tilde{R}$, with $R \in {\rm Spin}_{+}(1,3)$. One can
also prove
that any $R \in {\rm Spin}_{+}(1,3)$ can be written in the form $R =
\pm {\rm e}^{B}$
with $B \in \bigwedge^2(\mbox{${\sl  I\!\!R}$}^{1,3})$, and the
choice of the sign can
always be the positive one except when $R = -{\rm e}^B$ with $B^2 = 0$.
When $B$ is a timelike bivector ($B^2 > 0$) $R$ describes a boost,
while when $B$ is a spacelike bivector ($B^2 < 0$) $R$ describes
a spatial rotation.

We said that Clifford algebras are isomorphic to matrix algebras.
In the case of $\mbox{${\sl  I\!\!R}$}_{1,3}$ it is isomorphic to
${\cal M}(2,\mbox{${\sl  I\!\!H}$})$,
the algebra of $2 \times 2$ matrices over the quaternions. This
isomorphism defines representations of STA. One representation
is:
\begin{eqnarray}
\label{representation}
\Gamma_{0} \leftrightarrow  \left( \begin{array}{cc}
              1 & 0 \\
                 0 & -1
           \end{array} \right) ,
\Gamma_{1} \leftrightarrow  \left( \begin{array}{cc}
                  0 & \mbox{\boldmath $i$} \\
                 \mbox{\boldmath $i$} & 0
           \end{array} \right) , \nonumber \\
\label{eq.15}
\vspace{2ex}
\Gamma_{2} \leftrightarrow  \left( \begin{array}{cc}
                  0 & \mbox{\boldmath $j$} \\
                 \mbox{\boldmath $j$} & 0
           \end{array} \right) ,
\Gamma_{3} \leftrightarrow   \left( \begin{array}{cc}
                  0 & \mbox{\boldmath $k$} \\
                 \mbox{\boldmath $k$} & 0
           \end{array} \right) ,
\end{eqnarray}
where $\mbox{\boldmath $i$}^2 = \mbox{\boldmath $j$}^2 =
\mbox{\boldmath $k$}^2 = -1$ and $\mbox{\boldmath $i$}\mbox{\boldmath
$j$} =
\mbox{\boldmath $k$}$, $\mbox{\boldmath $j$}\mbox{\boldmath $k$} =
\mbox{\boldmath $i$}$, $\mbox{\boldmath $k$}\mbox{\boldmath $i$} =
\mbox{\boldmath $j$}$.

Finally, consider the idempotent $e = \frac{1}{2}(1 + \Gamma_0 )$.
$I_{1,3} = \mbox{${\sl  I\!\!R}$}_{1,3}e$ is a minimal left ideal. One can
easily prove that
\begin{equation}
\label{eq.16}
I_{1,3} = \mbox{${\sl  I\!\!R}$}_{1,3}e = \mbox{${\sl  I\!\!R}$}_{1,3}^{+}e ,
\end{equation}
which is an important result to be used later.

\subsection{Dirac Algebra}

Consider the vector space $\mbox{${\sl  I\!\!R}$}^{4,1}$ and a basis
$\{E_a\}$ ($a =
0,1,2,3,4$) such that $E_{1}^{2} = E_{2}^{2} = E_{3}^{2} =
E_{4}^{2} = -E_{0}^{2} = 1$. Let $\mbox{${\sl  I\!\!R}$}_{4,1}$ be
its Clifford
algebra and $i = E_0 E_1 E_2 E_3 E_4$ be the volume element.
Note that $i^2 = -1$, but differently from STA now we have
$E_a i = i E_a$ ($\forall a$), that is, $\mbox{${\sl  I\!\!R}$}
\oplus i\mbox{${\sl  I\!\!R}$}$ is the
center of $\mbox{${\sl  I\!\!R}$}_{4,1}$. The volume element $i$
plays therefore
the role of imaginary unity. Let us define
\begin{equation}
\label{eq.17}
\Gamma_{\mu} = E_{\mu} E_{4}  \, \, \, \, \, \, (\mu = 0,1,2,3) .
\end{equation}
One can easily see from the above map and with $i$ playing the
role of imaginary unity that $\mbox{${\sl  I\!\!R}$}_{4,1}$ is isomorphic to
the complexified STA:
\begin{equation}
\label{eq.18}
\mbox{${\sl  I\!\!R}$}_{4,1} \simeq \mbox{${\sl  I\!\!\!\!C}$}
\otimes \mbox{${\sl  I\!\!R}$}_{1,3} .
\end{equation}
Moreover, the even subalgebra of $\mbox{${\sl  I\!\!R}$}_{4,1}$ is isomorphic
to $\mbox{${\sl  I\!\!R}$}_{1,3}$:
\begin{equation}
\label{eq.19}
\mbox{${\sl  I\!\!R}$}_{4,1}^{+} \simeq \mbox{${\sl  I\!\!R}$}_{1,3} .
\end{equation}

The algebra $\mbox{${\sl  I\!\!R}$}_{4,1}$, or equivalently, the
complexified STA, is
called Dirac algebra. In fact, they are isomorphic to ${\cal
M}(4,\mbox{${\sl  I\!\!\!\!C}$})$
-- the algebra of $4 \times 4$ matrices over the complexes. One
representation (the standard one) of $\Gamma_{\mu}$ in eq.(\ref{eq.17}) is:
\begin{equation}
\label{eq.20}
\begin{array}{ll}
\Gamma_{0} \leftrightarrow \left( \begin{array}{cccc}
                1 & 0 & 0 & 0 \\
                0 & 1 & 0 & 0 \\
                0 & 0 & -1 & 0 \\
                0 & 0 & 0 & -1
              \end{array}  \right) , &
\Gamma_{1} \leftrightarrow \left ( \begin{array}{cccc}
                0 & 0 & 0 & -1 \\
                0 & 0 & -1 & 0 \\
                0 & 1 & 0 & 0 \\
                1 & 0 & 0 & 0
              \end{array}  \right) ,  \\
\vspace{2ex}
\Gamma_{2} \leftrightarrow \left( \begin{array}{cccc}
                0 & 0 & 0 & i \\
                0 & 0 & -i & 0 \\
                0 & -i & 0 & 0 \\
                i & 0 & 0 & 0
              \end{array}  \right) , &
\Gamma_{3} \leftrightarrow \left( \begin{array}{cccc}
                0 & 0 & -1 & 0 \\
                0 & 0 & 0 & 1 \\
                1 & 0 & 0 & 0 \\
                0 & -1 & 0 & 0
              \end{array}  \right) .
\end{array}
\end{equation}

Consider the idempotent $f = \frac{1}{2}(1 + \Gamma_0 )\frac{1}{2}
(1 + i \Gamma_{12}) = e \frac{1}{2}(1 + i \Gamma_{12})$, where
$e = \frac{1}{2}(1 + \Gamma_{0})$ is a primitive idempotent of
$\mbox{${\sl  I\!\!R}$}_{1,3}$. Then $I_{4,1} = \mbox{${\sl
I\!\!R}$}_{4,1}f$ is a minimal left
ideal, and one can show that
\begin{equation}
\label{eq.21}
I_{4,1} = \mbox{${\sl  I\!\!R}$}_{4,1}f \simeq \mbox{${\sl
I\!\!R}$}_{4,1}^{+}f .
\end{equation}

\subsection{Dirac-Hestenes Spinors}

Let us consider a Dirac spinor $\mid \! \! \Psi \rangle \in
\mbox{${\sl  I\!\!\!\!C}$}^4 $.
There is an obvious isomorphism between $\mbox{${\sl  I\!\!\!\!C}$}^4
$ and minimal left
ideals of ${\cal M}(4,\mbox{${\sl  I\!\!\!\!C}$})$, given by
\begin{equation}
\label{eq.22}
\mbox{${\sl  I\!\!\!\!C}$}^4 \ni \, \mid \! \! \Psi \rangle =
\left( \begin{array}{c}
       \psi_1 \\ \psi_2 \\ \psi_3 \\ \psi_4
       \end{array} \right)
\leftrightarrow \left( \begin{array}{cccc}
                \psi_1 & 0 & 0 & 0 \\
                \psi_2 & 0 & 0 & 0 \\
                \psi_3 & 0 & 0 & 0 \\
                \psi_4 & 0 & 0 & 0
              \end{array}  \right) = \Psi \in {\rm ideal} \, \,  {\rm
of } \, \,  {\cal M}(4,\mbox{${\sl  I\!\!\!\!C}$}).
\end{equation}

One can, of course, work with $\Psi$ instead of $\mid \! \! \Psi \rangle$,
and since ${\cal M}(4,\mbox{${\sl  I\!\!\!\!C}$})$ is a
representation of Dirac algebra
$\mbox{${\sl  I\!\!R}$}_{4,1} \simeq \mbox{${\sl  I\!\!\!\!C}$}
\otimes \mbox{${\sl  I\!\!R}$}_{1,3}$, one can work with
the corresponding ideal of Dirac algebra. Note that
\begin{equation}
\label{eq.23}
\left( \begin{array}{cccc}
                \psi_1 & 0 & 0 & 0 \\
                \psi_2 & 0 & 0 & 0 \\
                \psi_3 & 0 & 0 & 0 \\
                \psi_4 & 0 & 0 & 0
              \end{array}  \right) =
\underbrace{\left( \begin{array}{cccc}
                \psi_1 & \cdot & \cdot & \cdot \\
                \psi_2 & \cdot & \cdot & \cdot \\
                \psi_3 & \cdot & \cdot & \cdot \\
                \psi_4 & \cdot & \cdot & \cdot
              \end{array}  \right) }_{\in {\cal M}(4,{\sl C})}
\underbrace{\left( \begin{array}{cccc}
                1 & 0 & 0 & 0 \\
                0 & 0 & 0 & 0 \\
                0 & 0 & 0 & 0 \\
                0 & 0 & 0 & 0
              \end{array}  \right)}_{f} ,
\end{equation}
where $f$ is a matrix representation of the idempotent $f =
\frac{1}{2}(1 + \Gamma_0 )\frac{1}{2}(1 + i\Gamma_{12})$. One can
work therefore with the ideal $I_{4,1} = \mbox{${\sl
I\!\!R}$}_{4,1}f$ instead of
$\mbox{${\sl  I\!\!\!\!C}$}^4$. But the isomorphisms discussed in the
preceeding subsections
tell us that
\begin{equation}
\label{eq.24}
I_{4,1} = \mbox{${\sl  I\!\!R}$}_{4,1}f = (\mbox{${\sl
I\!\!\!\!C}$}\otimes\mbox{${\sl  I\!\!R}$}_{1,3})f \simeq
\mbox{${\sl  I\!\!R}$}_{4,1}^{+}f \simeq
\mbox{${\sl  I\!\!R}$}_{1,3}f = (\mbox{${\sl
I\!\!R}$}_{1,3}e)\frac{1}{2}
(1 + i\Gamma_{12}) .
\end{equation}
Note that in the last equality we have a minimal left ideal
$\mbox{${\sl  I\!\!R}$}_{1,3}e$ of STA. These equalities show that
all informations we
obtain from an element of the ideal $I_{4,1}$ can be obtained from an
element of the ideal $I_{1,3} = \mbox{${\sl  I\!\!R}$}_{1,3}e$.
Moreover, we have that
\begin{equation}
\label{eq.25}
i f = \Gamma_{21} f .
\end{equation}
These results mean that we can work with the ideal $\mbox{${\sl
I\!\!R}$}_{1,3}e$ once
we identify $\Gamma_{21}$ as playing in STA the role of the imaginary
unity $i$.

Now, we saw that
\begin{equation}
\label{eq.26}
\mbox{${\sl  I\!\!R}$}_{1,3}e = \mbox{${\sl  I\!\!R}$}_{1,3}^{+}e .
\end{equation}
What the idempotent makes is ``kill" redundant degrees of freedom.
Since ${\rm dim} \mbox{${\sl  I\!\!R}$}_{1,3}^{+} = 8$ we can work
with $\mbox{${\sl  I\!\!R}$}_{1,3}^{+}$
instead of $\mbox{${\sl  I\!\!R}$}_{1,3}e$ (this is not the case for
$\mbox{${\sl  I\!\!R}$}_{1,3}$
or $\mbox{${\sl  I\!\!\!\!C}$}\otimes\mbox{${\sl  I\!\!R}$}_{1,3}$
since ${\rm dim}\mbox{${\sl  I\!\!R}$}_{1,3} = 16$ and
${\rm dim}(\mbox{${\sl  I\!\!\!\!C}$}\otimes\mbox{${\sl
I\!\!R}$}_{1,3}) = 32$). We established therefore
the isomorphism
\begin{equation}
\label{eq.27}
\mbox{${\sl  I\!\!\!\!C}$}^4 \simeq \mbox{${\sl  I\!\!R}$}_{1,3}^{+} .
\end{equation}
The representative of the Dirac spinor $\mid \! \! \Psi \rangle$ in
$I\!\!R_{4,1}$ is
$\Psi$, which is related to $\psi \in \mbox{${\sl  I\!\!R}$}_{1,3}^{+}$ by
\begin{equation}
\label{eq.28}
\Psi = \psi \frac{1}{2}(1 + \Gamma_0 )\frac{1}{2}
(1 + i\Gamma_{12})  .
\end{equation}
Such $\psi$ will be called Dirac-Hestenes spinor. Its (standard)
matrix representation is:
\begin{equation}
\label{eq.28a}
\psi \leftrightarrow  \left( \begin{array}{cccc}
             \psi_1 & -\psi_2^* & \psi_3 & \psi_4^* \\
             \psi_2 & \psi_1^*  & \psi_4 & -\psi_3^* \\
             \psi_3 & \psi_4^* & \psi_1  & -\psi_2^* \\
             \psi_4 & -\psi_3^* & \psi_2 & \psi_1^*
               \end{array} \right) .
\end{equation}

Now, $\psi \in {\sl I \!\! R}_{1,3}^{+}$, so that its general form is
\begin{eqnarray}
\label{eq.28b}
\psi = && a + a_{01}\Gamma_{01} + a_{02}\Gamma_{02}
+ a_{03}\Gamma_{03} \nonumber \\
&&+ a_{12}\Gamma_{12} + a_{13}\Gamma_{13} + a_{23}\Gamma_{23}
+ a_{0123}\Gamma_{5} .
\end{eqnarray}
{}From the representation (\ref{representation}) we have
\begin{eqnarray}
\label{eq.28c}
&&\Gamma_{12} = \mbox{\boldmath $k$} 1, \quad
\Gamma_{31} = \mbox{\boldmath $j$} 1, \quad
\Gamma_{23} = \mbox{\boldmath $i$} 1, \nonumber \\
&&\Gamma_{01} = -\Gamma_{5}\mbox{\boldmath $i$} 1, \quad
\Gamma_{02} = -\Gamma_{5}\mbox{\boldmath $j$} 1, \quad
\Gamma_{03} = -\Gamma_{5}\mbox{\boldmath $k$} 1,
\end{eqnarray}
where $1$ is the identity $2 \times 2$ matrix and $\Gamma_5$ commutes
with the quaternion units $\mbox{\boldmath $i$}$,  $\mbox{\boldmath
$j$}$, $\mbox{\boldmath $k$}$. $\psi$ can therefore be represented by
a biquaternion number, that is, by $A + \Gamma_5 B$ with $A$ and $B$
quaternions, or, since $\Gamma_5$ commutes with $\mbox{\boldmath
$i$}$,  $\mbox{\boldmath $j$}$, $\mbox{\boldmath $k$}$ and
$\Gamma_5^2 = -1$, by a quaternion over the complexes.

One important comment that must be made here concerns the
transformation law of a Dirac-Hestenes spinor. In order to obtain a
correct transformation law under the action of the Lorentz group, we
must be more precise \cite{equi} and say that a Dirac-Hestenes spinor
is an element of the quotient set ${\sl I \!\! R}^{+}_{1,3}
/ {\cal R}$ such that given the orthonormal basis $\Sigma$ and
$\dot{\Sigma}$ of ${\sl I \!\! R}^{1,3} \subset {\sl I \!\!
R}_{1,3}$, $\psi_{\scriptscriptstyle \Sigma} \in {\sl I \!\!
R}_{1,3}^{+}$, $\psi_{\dot{\scriptscriptstyle \Sigma}} \in {\sl I \!\!
R}_{1,3}^{+}$, then $\psi_{\scriptscriptstyle \Sigma} \sim
\psi_{\dot{\scriptscriptstyle \Sigma}}$ (mod ${\cal R}$) if and only
if $\psi_{\dot{\scriptscriptstyle \Sigma}} =
\psi_{\scriptscriptstyle \Sigma} U^{-1}$ with $\dot{\Sigma} =
{\cal L}(\Sigma) = U\Sigma U^{-1}$, $U \in {\rm Spin}_{+}(1,3)$,
${\cal L} \in {\rm SO}_{+}(1,3)$ and ${\cal H}(U) = {\cal L}$ where
${\cal H}$ is the double covering ${\cal H}: {\rm Spin}_{+}(1,3)
\rightarrow {\rm SO}_{+}(1,3)$. We say that
$\psi_{\scriptscriptstyle \Sigma}$ is the representative of the
Dirac-Hestenes spinor in the orthonormal basis $\Sigma$, and when no
confusion arises we write only $\psi$ instead of
$\psi_{\scriptscriptstyle \Sigma}$.

Suppose now that $\psi$ is non-singular, i.e., $\psi\tilde{\psi} \neq 0$.
Since $\psi \in \mbox{${\sl  I\!\!R}$}_{1,3}^{+}$ we have
\begin{equation}
\label{eq.29}
\psi\tilde{\psi} = \sigma + \Gamma_5 \omega ,
\end{equation}
where $\sigma$ and $\omega$ are scalars. Define quantities $\rho =
\sqrt{\sigma^2 + \omega^2}$ and $\tan{\beta} = \omega / \sigma$. Then
$\psi$ can be written as
\begin{equation}
\label{eq.30}
\psi = \sqrt{\rho}{\rm e}^{\Gamma_5 \beta /2} R ,
\end{equation}
where $R \in {\rm Spin}_{+}(1,3)$, $\rho \in \mbox{${\sl
I\!\!R}$}_{+}$ and $0 \leq \beta < 2\pi$.
This is the canonical decomposition of Dirac-Hestenes spinors in
terms of a Lorentz rotation $R$, a dilation $\rho$ and a duality
transformation by an angle $\beta$ (called Yvon-Takabayasi angle).
This is a remarkable result since it generalizes the polar
decomposition of a complex number for the case of a biquaternion,
and has a clear geometrical interpretation.

Finally, let us write Dirac equation in terms of $\psi$.
Let $(M,g,D)$ be Minkowski spacetime, where $(M,g)$ is a four
dimensional time oriented and space oriented Lorentzian manifold with
$g$ a Lotentzain metric with signature $(1,3)$ and $D$ is its
Levi-Civita connection. The tangent space at $x \in M$ is
$T_x M \simeq {\sl I \!\! R}^{1,3}$ and the tangent bundle is
$T(M) = \cup_{x\in M} T_x M$. We represent by $\bigwedge(M) =
\cup_{x\in M}\bigwedge(T_x M)$ the Cartan bundle of multivectors
and by ${\cal C}\ell (M) = \cup_{x\in M}{\cal C}\ell (T_x M)$ the
Clifford bundle of multivectors, where ${\cal C}\ell (T_x M) \simeq
{\sl I \!\! R}_{1,3}$. Dirac-Hestenes spinor fields are sections of
the so called Spin-Clifford bundle ${\cal C}\ell_{{\rm
Spin}_{+}(1,3)}(M) \simeq {\cal C}\ell (M) / {\cal R}$, where ${\cal
R}$ is the equivalence relation defined above. This means that given
an orthonormal basis $\Sigma = \{\gamma_{\mu}\}$, where
$\gamma_{\mu}  \in \sec T(M) \in {\rm sec}{\cal C}\ell(M)$, the
representative of a Dirac-Hestenes spinor field in the basis $\Sigma$
is an even section of ${\cal C}\ell (M)$. Note that for each $x \in
M$ we have $\gamma_{\mu} \simeq \Gamma_{\mu}$.
Introducing $\{\gamma^{\mu}\}$ as the reciprocal basis, i.e.,
$\gamma^{\mu}\cdot\gamma_{\nu}=\delta^{\mu}_{\nu}$.
We have $\gamma^{\mu} \simeq \Gamma^{\mu}$ for each $x \in M$.
With the above remarks, and using some previous results,
namely $\mid \! \! \Psi \rangle = \psi f$, $\gamma_{21}f = if$ and
$\gamma_0 f = f$,
the (free) Dirac equation $i\partial \mid \! \! \Psi(x) \rangle = (mc/\hbar)
\mid \! \! \Psi(x) \rangle $, where
$\mid \!\! \Psi(x)\rangle$ is a Dirac spinor
field \cite{Choquet}, is written in STA as \cite{Oli}
\begin{equation}
\label{eq.31}
\partial \psi \gamma_{21} = \frac{mc}{\hbar} \psi \gamma_0  ,
\end{equation}
which we call Dirac-Hestenes equation. $\partial$ is the Dirac
operator, which in an orthonormal coordinate basis is $\partial =
\gamma^{\mu}\frac{\partial}{\partial x^\mu}$.

It is interesting to observe that $\gamma_{21}$ appears on the right
of $\psi$ in eq.(\ref{eq.31}). The reason is the equation
$\gamma_{21} f = i f$, from which we have $\psi\gamma_{21} f = \psi i
f = i \psi f = i \mid \! \psi \rangle$. If we represent $\psi$ by
means of quaternions, a quaternion unit has to be introduced
multiplying $\psi$ on the right. This is the reason why de Leo and
Rotelli \cite{Leo}  had to introduced left and right acting elements
in their quaternionic version of quantum mechanics.

\section{Spinorial Representation of Maxwell Equations}

The spinorial representation of Maxwell equations we shall give
in this section is based on the following theorem:

\vspace{1ex}
\noindent {\bf Theorem:} Any electromagnetic field $F \in
{\rm sec}\bigwedge^2(M) \in {\rm sec}{\cal C}\ell(M)$ can be written
in the form
\begin{equation}
\label{eq.32}
F = \psi \gamma_{21} \tilde{\psi}  ,
\end{equation}
where $\psi$ is a Dirac-Hestenes spinor field.

\vspace{1ex}

The proof of this theorem can be divided in three steps: (i) $F^2 \neq 0$;
(ii) $F^2 = 0$, $F \neq 0$; (ii) $F^2 = 0$, $F = 0$. In the first case
the proof is based on a theorem by Rainich \cite{Rai},
and reconsidered by Misner and Wheeler \cite{MW}.

\vspace{1ex}

\noindent {\bf Theorem (Rainich-Misner-Wheeler):} Let an extremal
field be an electromagnetic field for which the electric [magnetic]
field vanishes and the magnetic [electric] field is parallel to a
given spatial direction. Then, at any point of spacetime, any
non-null ($F^2 \neq 0$) electromagnetic field $F$ can be transformed
in an extremal field by means of a Lorentz transformation and a
duality transformation.

\vspace{1ex}

An easy proof of the theorem of Rainich-Misner-Wheeler can be found
in \cite{IJTP}. Now consider a Lorentz transformation $F \mapsto
F^{\prime} = L F \tilde{L}$, and a duality transformation $F^{\prime}
\mapsto F^{\prime\prime} = {\rm e}^{\gamma_5 \alpha}F^{\prime}$.
According to the theorem of RMW the electromagnetic field $F^{\prime\prime}$
is extremal; let us suppose it of magnetic type along the $\vec{k}$
direction, that is: $F^{\prime\prime} = \rho \gamma_{21}$, where
$\rho$ is the extremal field intensity. We have therefore that
$\rho \gamma_{21} = {\rm e}^{\gamma_5 \alpha} L F \tilde{L}$. Let us
define $R = \tilde{L}$ and $\beta = -\alpha$; then we have that
\begin{equation}
\label{eq.33}
F = \psi \gamma_{21} \tilde{\psi} ,
\end{equation}
where
\begin{equation}
\label{eq.34}
\psi = \sqrt{\rho} {\rm e}^{\gamma_{5}\beta /2} R ,
\end{equation}
which we recognize as the canonical decomposition of Dirac-Hestenes
spinor.

In order to prove our theorem for case (ii) we observe that since
$F^2 = 0$ we have $\vec{E}\cdot\vec{H} = 0$ and $\vec{E}^2 = \vec{H}^2$;
we can make therefore a spatial rotation ${\cal R}$ such that $E_{1}^{\prime}
= H_{1}^{\prime} = 0$ and $H_{3}^{\prime} = \pm E_{2}^{\prime} = \eta_1$
and $H_{2}^{\prime} = \pm E_{3}^{\prime} = \eta_2$; then for $F^{\prime}
= (1/2) (F^{\prime})^{\mu\nu}\gamma_{\mu\nu}$ we have $F^{\prime} =
(\eta_1 + \gamma_5 \eta_2)(1/2)(1 \pm \gamma_{01})\gamma_{21}$. If we
take $R = \tilde{\cal R}$ we have for $F = R F^{\prime} \tilde{R}$ that
$F = (\eta_1 + \gamma_5 \eta_2)R (1/2)(1 \pm \gamma_{01})\gamma_{21}
\tilde{R}$. Remember that $(1/2)(1 \pm \gamma_{01})$ is an idempotent;
defining $\eta_1 = \eta \cos{\varphi}$ and $\eta_2 = \eta \sin{\varphi}$
it follows that
\begin{equation}
\label{eq.35}
F = \psi_{\scriptscriptstyle M}\gamma_{21}
\tilde{\psi}_{\scriptscriptstyle M}  ,
\end{equation}
where
\begin{equation}
\label{eq.36}
\psi_{\scriptscriptstyle M} = \sqrt{\eta}{\rm e}^{\gamma_5 \varphi /2}
R \frac{1}{2}(1 \pm \gamma_{01}) = \psi \frac{1}{2}( 1 \pm \gamma_{01}) ,
\end{equation}
which proves our assertion in this case. $\psi_{\scriptscriptstyle M}$
is a particular type of Dirac-Hestenes spinor known as Majorana
spinor \cite{Lounesto}.

Now for case (iii) ($F = 0$) we note that $\psi\gamma_{21}\tilde{\psi} =
- \psi\gamma_{21}\tilde{\psi} = \gamma_5 \psi\gamma_{21}\tilde{\psi}\gamma_5
= \gamma_5 \psi\gamma_{21}\gamma_{21}\gamma_{12}\tilde{\psi}\gamma_5 $ is
satisfied for $\psi = \pm \gamma_5 \psi \gamma_{21}$. It follows
therefore that
\begin{equation}
\label{eq.37}
F = \psi_{\scriptscriptstyle W}\gamma_{21}\tilde{\psi}_{\scriptscriptstyle W}
= 0 ,
\end{equation}
where
\begin{equation}
\label{eq.38}
\psi_{\scriptscriptstyle W} = \frac{1}{2}(\psi \pm \gamma_5 \psi
\gamma_{21})  .
\end{equation}
This particular kind of Dirac-Hestenes spinor is called a Weyl
spinor \cite{Lounesto}. We have now proved our theorem. In terms of
usual Dirac spinor fields eq.(\ref{eq.32}) gives
\begin{equation}
\label{eq.38.0}
F_{\mu\nu} = \langle \Psi \! \!
\mid \frac{i}{2}[\gamma_{\mu},\gamma_{\nu}]
\mid \! \! \Psi \rangle ,
\end{equation}
where $\gamma_{\mu}$ are
Dirac matrices, that is, the matrix representation of vectors $\gamma_{\mu}$
as in eq.(\ref{eq.20}),  $\mid \! \! \Psi \rangle$ is the Dirac
spinor field and $\langle \Psi \! \! \mid$ its Dirac adjoint spinor
field \cite{Choquet}.

\vspace{1ex}

\noindent {\bf Remark 1:} The Faraday bivector $F$ is given by a
quadractic expression in terms of the Dirac-Hestenes spinor according
to eq.(\ref{eq.32}). In this way we speak of the Faraday bivector as
the ``Dirac square'' of the Dirac-Hestenes spinor, since
eq.(\ref{eq.32}) is a particular kind of square involving the
bivector $\gamma_{21}$ which plays the
role of an unity, in this case an unity of extremal field. Similarly,
we speak of the Dirac-Hestenes spinor as the ``Dirac square root'' of
the Faraday bivector.

\vspace{1ex}
\noindent {\bf Remark 2:} It remains, of course, the question of
the constants in eq.(\ref{eq.32}) since the units of $F$ are charge
$\times$ (length)$^{-2}$ and the units of $\psi$ are (length)$^{-3/2}$.
We have now to introduce two postulates: firstly, we suppose that there
is a natural unit $F_0$ of electromagnetic field intensity; secondly,
we suppose that there is a natural unit $e_0$ of electric charge. In this
way we have that
\begin{equation}
\label{eq.38a}
F = \left( \frac{e_0^3}{F_0} \right)^{\frac{1}{2}} \psi \gamma_{21}
\tilde{\psi} .
\end{equation}
Note that $(e_0 / F_0 )^{1/2}$ has unit of length. One of the above postulates
can be replaced by the one that there is a natural unit $L_0$ of length,
which gives $F_0 = e_0 / L_0^2 $ and
\begin{equation}
\label{eq.38b}
F = e_0 L_0 \psi \gamma_{21} \tilde{\gamma} .
\end{equation}
In terms of some physical constants, a combination of
constants we need can be
\begin{equation}
\label{eq.39}
F = \frac{e\hbar}{2mc}\psi\gamma_{21}\tilde{\psi} ,
\end{equation}
which gives correct units for $F$ and $\psi$. In this expression
the symbols have their usual meaning, that is: $e$ is the elementary
electric charge, $\hbar$ is the Planck constant, $m$ is the electron
mass and $c$ the velocity of light in vacuum. In what follows we
will work, to simplify the notation,  with eq.(\ref{eq.32}) instead
of eq.(\ref{eq.39}).

\vspace{1ex}

The idea now is to use $F = \psi\gamma_{21}\tilde{\psi}$ in Maxwell
equations and obtain from it an equivalent equation for $\psi$.
Maxwell equations using STA are written as an unique equation,
namely \cite{He66,Maia}
\begin{equation}
\label{eq.40}
\partial F = {\cal J} ,
\end{equation}
where $\partial = \gamma^{\mu}\partial_{\mu}$ is the Dirac operator
and ${\cal J}$ is the electromagnetic current (an electric current
$J_e$ plus a magnetic monopolar current $\gamma_{5} J_m$ in the general
case). If we use eq.(\ref{eq.32}) in Maxwell equation (\ref{eq.40})
we obtain
\begin{equation}
\label{eq.41}
\partial(\psi\gamma_{21}\tilde{\psi}) = \gamma^{\mu}\partial_{\mu}
(\psi\gamma_{21}\tilde{\psi}) =
\gamma^{\mu}(\partial_{\mu}\psi
\gamma_{21}\tilde{\psi} + \psi \gamma_{21}\partial_{\mu}\tilde{\psi}) =
{\cal J}  .
\end{equation}
But $\psi\gamma_{21}\partial\tilde{\psi} = -(\partial_{\mu}\psi
\gamma_{21}\tilde{\psi})\, \tilde{}$, and since reversion does not
change the sign of scalars and of pseudo-scalars (4-vectors), we have that
\begin{equation}
\label{eq.42}
2\gamma^{\mu}\langle\partial_{\mu}\psi\gamma_{21}\tilde{\psi}\rangle_2
= {\cal J} .
\end{equation}
There is a more convenient way of rewriting the above equation. Note that
\begin{equation}
\label{eq.43}
\gamma^{\mu}\langle\partial_{\mu}\psi\gamma_{21}\tilde{\psi}\rangle_2 =
\partial\psi\gamma_{21}\tilde{\psi} -
\gamma^{\mu}\langle\partial_{\mu}\psi\gamma_{21}\tilde{\psi}\rangle_0  -
\gamma^{\mu}\langle\partial_{\mu}\psi\gamma_{21}\tilde{\psi}\rangle_4 ,
\end{equation}
and if we define the vectors
\begin{equation}
\label{eq.44}
j = \gamma^{\mu}\langle\partial_{\mu}\psi\gamma_{21}\tilde{\psi}\rangle_0 ,
\end{equation}
\begin{equation}
\label{eq.45}
g = \gamma^{\mu}\langle\partial_{\mu}\psi\gamma_5
\gamma_{21}\tilde{\psi}\rangle_0  ,
\end{equation}
we can rewrite eq.(\ref{eq.42}) as
\begin{equation}
\label{eq.46}
\partial\psi\gamma_{21}\tilde{\psi} =
\left[ \frac{1}{2}{\cal J} + \left( j + \gamma_5 g \right) \right]  .
\end{equation}
If correct units have been used, that is, if we used eq.(\ref{eq.39}),
then instead of $(1/2){\cal J}$ we would obtained $(mc/e\hbar){\cal J}$.
Eq.(\ref{eq.46}) is the spinorial representation of Maxwell equations
we were looking for. In the case where $\psi$ is non-singular (which
corresponds to non-null electromagnetic fields) we have
\begin{equation}
\label{eq.47}
\partial\psi\gamma_{21} = \frac{{\rm e}^{\gamma_5 \beta}}{\rho}
\left[ \frac{1}{2}{\cal J} +
\left( j + \gamma_5 g \right) \right] \psi  .
\end{equation}
Eq.(\ref{eq.47}) has been proved \cite{IJTP} to be equivalent to
the spinorial representation of Maxwell equations obtained originally
by Campolattaro \cite{Cam90} in terms of the
usual covariant Dirac spinor.

\section{Is There Any Relationship Between Maxwell and Dirac
Equations?}

The spinorial eq.(\ref{eq.47}) that represents Maxwell ones, as
written in that form, does not appear to have any relationship with
Dirac equation (\ref{eq.31}). However, we shall make some modifications
on it in such a way to put it in a form that suggests a very interesting
and intriguing relationship between them, and consequently between
electromagnetism and quantum mechanics.

Since $\psi$ is supposed to be non-singular ($F$ non-null) we can use
the canonical decomposition (\ref{eq.30}) of $\psi$ and write
\begin{equation}
\label{eq.48}
\partial_{\mu}\psi = \frac{1}{2}\left( \partial_{\mu}\ln{\rho} +
\gamma_5 \partial_{\mu}\beta + \Omega_{\mu} \right)\psi ,
\end{equation}
where we defined
\begin{equation}
\label{eq.49}
\Omega_{\mu} = 2(\partial_{\mu}R)\tilde{R}  .
\end{equation}
Using this expression for $\partial_{\mu}\psi$ into the definitions
of the vectors $j$ and $g$ (eqs.(\ref{eq.44},\ref{eq.45})) we obtain
that
\begin{equation}
\label{eq.50}
j = \gamma^{\mu}(\Omega_{\mu}\cdot S)\rho\cos{\beta} +
\gamma_{\mu}[\Omega_{\mu}\cdot(\gamma_5 S)]\rho\sin{\beta} ,
\end{equation}
\begin{equation}
\label{eq.51}
g = \gamma^{\mu}[(\Omega_{\mu}\cdot (\gamma_5 S)]\rho\cos{\beta} -
\gamma_{\mu}(\Omega_{\mu}\cdot S)\rho\sin{\beta} ,
\end{equation}
where we defined the bivector $S$ by
\begin{equation}
\label{eq.52}
S = \frac{1}{2}\psi\gamma_{21}\psi^{-1} = \frac{1}{2}R\gamma_{21}\tilde{R}  .
\end{equation}
A more convenient expression can be written. Let $v$ be given by
$\rho v = J = \psi \gamma_0 \tilde{\psi}$, and $v_{\mu} = v\cdot\gamma_{\mu}$.
Define the bivector $\Omega = v^{\mu}\Omega_{\mu}$ and the scalars
$\Lambda$ and $K$ by
\begin{equation}
\label{eq.53}
\Lambda = \Omega \cdot S ,
\end{equation}
\begin{equation}
\label{eq.54}
K = \Omega \cdot (\gamma_5 S) .
\end{equation}
Using these definitions we have that
\begin{equation}
\label{eq.55}
\Omega_{\mu}\cdot S = \Lambda v_{\mu} ,
\end{equation}
\begin{equation}
\label{eq.56}
\Omega_{\mu}\cdot(\gamma_5 S) = K v_{\mu}  ,
\end{equation}
and for the vectors $j$ and $g$:
\begin{equation}
\label{eq.57}
j = \Lambda v \rho\cos{\beta} + K v \rho\sin{\beta} = \lambda \rho v ,
\end{equation}
\begin{equation}
\label{eq.58}
g = K v \rho\cos{\beta} - \Lambda v \rho \sin{\beta} = \kappa \rho v ,
\end{equation}
where we defined
\begin{equation}
\label{eq.59}
\lambda = \Lambda \cos{\beta} + K \sin{\beta} ,
\end{equation}
\begin{equation}
\label{eq.60}
\kappa = K \cos{\beta} - \Lambda \sin{\beta}  .
\end{equation}
The spinorial representation of Maxwell equations are written
now as
\begin{equation}
\label{eq.61}
\partial\psi\gamma_{21} = \frac{{\rm e}^{\gamma_5 \beta}}{2\rho}{\cal J}\psi +
\lambda\psi\gamma_0 + \gamma_5 \kappa\psi\gamma_0  .
\end{equation}
If ${\cal J} = 0$ (free case) we have that
\begin{equation}
\label{eq.62}
\partial\psi\gamma_{21} =
\lambda\psi\gamma_0 + \gamma_5 \kappa\psi\gamma_0  ,
\end{equation}
which is very similar to Dirac equation (\ref{eq.31}).

In order to go a step further into the relationship between those
equations, we remember that the electromagnetic field has six
degrees of freedom, while a Dirac-Hestenes spinor field has
eight degrees of freedom; we are free therefore to impose two
constraints on $\psi$ if it is to represent an electromagnetic
field \cite{Cam90}. We choose these two constraints as
\begin{equation}
\label{eq.63}
\partial \cdot j = 0 \, \, \, \, \, \, {\rm and} \, \, \, \, \, \,
\partial \cdot g = 0 .
\end{equation}
Using eqs.(\ref{eq.57},\ref{eq.58}) these two constraints become\
\begin{equation}
\label{eq.64}
\partial \cdot j = \rho \dot{\lambda} + \lambda\partial\cdot J = 0 ,
\end{equation}
\begin{equation}
\label{eq.65}
\partial \cdot g = \rho \dot{\kappa} + \kappa\partial\cdot J = 0 ,
\end{equation}
where $J = \rho v$ and $\dot{\lambda} = (v\cdot\partial)\lambda$,
$\dot{\kappa} = (v\cdot\partial)\kappa$. These conditions imply that
\begin{equation}
\label{eq.66}
\kappa\dot{\lambda} = \lambda\dot{\kappa}  ,
\end{equation}
which gives ($\lambda \neq 0$):
\begin{equation}
\label{eq.67}
\frac{\kappa}{\lambda} = {\rm const} = -\tan{\beta_0} ,
\end{equation}
or from eqs.(\ref{eq.59},\ref{eq.60}):
\begin{equation}
\label{eq.68}
\frac{K}{\Lambda} = \tan{(\beta - \beta_0)} .
\end{equation}

Now we observe that $\beta$ is the angle of the duality rotation
from $F$ to $F^{\prime} = {\rm e}^{\gamma_5 \beta}F$. If we perform
another duality rotation by $\beta_0$ we have $F \mapsto {\rm e}^{\gamma_5
(\beta + \beta_0)}F$, and for the Yvon-Takabayasi angle $\beta \mapsto
\beta + \beta_0$. If we work therefore with an electromagnetic field
duality rotated by an additional angle $\beta_0$, the above
relationship becomes
\begin{equation}
\label{eq.69}
\frac{K}{\Lambda} = \tan{\beta} .
\end{equation}
This is, of course, just a way to say that we can choose the constant
$\beta_0$ in eq.(\ref{eq.67}) to be zero. Now,
this expression gives
\begin{equation}
\label{eq.70}
\lambda = \Lambda \cos{\beta} + \Lambda \tan{\beta}\sin{\beta} =
\frac{\Lambda}{\cos{\beta}}  ,
\end{equation}
\begin{equation}
\label{eq.71}
\kappa = \Lambda \tan{\beta}\cos{\beta} - \Lambda \sin{\beta} = 0 ,
\end{equation}
and the spinorial representation (\ref{eq.62}) of the free Maxwell
equations becomes
\begin{equation}
\label{eq.72}
\partial\psi\gamma_{21} = \lambda\psi\gamma_0  .
\end{equation}

Note that $\lambda$ is such that
\begin{equation}
\label{eq.73}
\rho\dot{\lambda}  = -\lambda \partial\cdot J .
\end{equation}
The current $J = \psi\gamma_0\tilde{\psi}$ is not conserved unless
$\lambda$ is constant. If we suppose also that
\begin{equation}
\label{eq.74}
\partial \cdot J = 0
\end{equation}
we must have
\begin{equation}
\label{eq.75}
\lambda = {\rm const.}
\end{equation}

Now, throughout these calculations we have assumed $\hbar = c = 1$.
We observe that in eq.(\ref{eq.72}) $\lambda$ has the units of
(length)$^{-1}$, and if we introduce the constants $\hbar$ and $c$
we have to introduce another constant with unit of mass. If we
denote this constant by $m$ such that
\begin{equation}
\label{eq.76}
\lambda = \frac{mc}{\hbar} ,
\end{equation}
then eq.(\ref{eq.72}) assumes a form which is identical to
Dirac equation:
\begin{equation}
\label{eq.77}
\partial\psi\gamma_{21} = \frac{mc}{\hbar}\psi\gamma_0  .
\end{equation}

It is true that we didn't proved that eq.(\ref{eq.77}) is really
Dirac equation since the constant $m$ has to be identified in
this case with the electron's mass. However, we shall make some
remarks concerning this identification
which are very interesting and intriguing. First, if in
analogy to eq.(\ref{eq.76}) we write
\begin{equation}
\label{eq.77a}
\Lambda = \frac{Mc}{\hbar},
\end{equation}
then eq.(\ref{eq.70}) reads
\begin{equation}
\label{eq.78}
m = \frac{M}{\cos{\beta}}  ,
\end{equation}
or
\begin{equation}
\label{eq.79}
M = m\cos{\beta} = \frac{m}{\sqrt{1 + \omega^2 / \sigma^2}} ,
\end{equation}
where $\sigma$ and $\omega$ are the invariants of Dirac theory and
given by eq.(\ref{eq.29}). If $m$ is constant, the above expression
defines a variable mass $M$.

On the other hand, de Broglie introduced
in his interpretation of quantum mechanics a variable mass $M$ related
to the constant one $m$ by de Broglie-Vigier formula \cite{dB1} $M =
m\sqrt{ 1 +
\omega^2 / \sigma^2}$, which is very similar the our one. However,
the difference in these formulas are unimportant for the free case
where we have for the plane wave solutions that \cite{He90}
$\cos{\beta} = \pm 1$.
Another interesting fact comes from eq.(\ref{eq.53}). If we write
$\psi = \psi_0 {\rm exp}(\gamma_{21}\omega t)$, where $\psi_0$ is a
constant spinor (which is the case again for plane waves), then
eq.(\ref{eq.53}) gives $\Lambda = \omega$, or, introducing the
constants $\hbar$ and $c$:
\begin{equation}
\label{eq.80}
Mc^2 = \hbar \omega .
\end{equation}
The variable mass $M$ now appears to be related to energy of
some ``internal'' vibration, and eq.(\ref{eq.80}) is just another
formula of de Broglie \cite{dB2}, who suggested that mass is related to
the frequency of an internal clock supposed associated to a
particle.

We can better understand the meaning of eq.(\ref{eq.53}) after some
additional manipulations. We have
\begin{equation}
\label{eq.80a}
\Lambda = \Omega \cdot S = v^{\mu}\langle \Omega_{\mu}\frac{1}{2}
\psi\gamma_{21}\psi^{-1} \rangle_0 ,
\end{equation}
and using eq.(\ref{eq.48}):
\begin{equation}
\label{eq.80b}
\Lambda = v^{\mu}\langle \partial_{\mu}\gamma_{21}\psi^{-1}\rangle_0 -
v^{\mu}(\partial_{\mu}\ln{\rho})\langle \psi\gamma_{21}\psi^{-1}
\rangle_0 - v^{\mu}\partial_{\mu}\beta\langle \gamma_5 \psi \gamma_{21}
\psi^{-1} \rangle_{0}  .
\end{equation}
The second and third terms on the RHS vanish because they are 2-vectors;
then, using  $\psi^{-1} = \tilde{\psi}(\psi\tilde{\psi})^{-1}$,
\begin{eqnarray}
&{}&\Lambda = v^{\mu}\langle \partial_{\mu}\psi \gamma_{21} \gamma_0
\tilde{\psi} (\psi\tilde{\psi})^{-1}\psi\gamma_0 \tilde{\psi}
(\psi\tilde{\psi})^{-1}
\rangle_0 = \nonumber \\
\label{eq.80c}
&{}& = v^{\mu}\langle \partial_{\mu}\psi \gamma_{21}\gamma_0 \tilde{\psi}
\frac{{\rm e}^{-\gamma_5 \beta}}{\rho} \rho v \frac{{\rm
e}^{-\gamma_5 \beta}}{\rho} \rangle_0 =
v^{\mu} \frac{1}{\rho} v^{\nu}\langle \partial_{\mu}\psi \gamma_{21}
\gamma_0 \tilde{\psi}\gamma_{\nu} \rangle_0 .
\end{eqnarray}
But one can show \cite{He90} that the energy-momentum tensor in Dirac theory
(Tetrode tensor) in terms of Dirac-Hestenes spinor is given by
\begin{equation}
\label{eq.80d}
T_{\mu\nu} = \langle \partial_{\mu}\psi \gamma_{21}\gamma_0 \tilde{\psi}
\gamma_{\nu} \rangle_0 .
\end{equation}
Since
\begin{equation}
\label{eq.80e}
v^{\nu}T_{\mu\nu} = \rho p_{\mu}
\end{equation}
it follows, with correct units, the well-known equation
\begin{equation}
\label{eq.80f}
M c = v \cdot p  .
\end{equation}
We remember that for plane waves $\cos{\beta} = \pm 1$, which implies from
eq.(\ref{eq.79}) that $p = \pm mcv$, with the plus sign corresponding
to the positive energy
solution and the minus one to the negative energy solution, and
which enable us to look as Feynman and Stueckelberg to
electron and positron as particles with opposite momenta.

\section{Other Possible Interpretations}

We can go on in our analysis and look to this problem in a
different way. Consider Dirac equation, now with an
external electromagnetic potential (introduced only for
convenience):
\begin{equation}
\label{eq.81}
\partial\phi\gamma_{21} = m\phi\gamma_0 + eA\phi  .
\end{equation}
Let us write the Dirac-Hestenes spinor field $\phi$ in the
following form
\begin{equation}
\label{eq.82}
\phi = \phi_0 {\rm e}^{-\gamma_{21}\chi} ,
\end{equation}
which is always possible, with $\phi_0 = \sqrt{\rho}{\rm e}^{\gamma_5
\beta /2}R_0$. If we use this expression in Dirac equation we obtain
\begin{equation}
\label{eq.83}
\partial\phi_0 \gamma_{21} + \partial\chi \phi_0 = m\phi_0 \gamma_0
+ eA\phi_0 .
\end{equation}
Since $J = \rho v = \phi\gamma_0 \tilde{\phi} = \phi_0 \gamma_0
\tilde{\phi}_0$, we can also write
\begin{equation}
\label{eq.84}
\partial\phi_0 \gamma_{21} \phi_{0}^{-1} + \partial\chi =
m{\rm e}^{\gamma_5 \beta}v + eA .
\end{equation}
This equation has vector and pseudo-vector (3-vector) parts, which are
\begin{equation}
\label{eq.85}
\langle \partial\phi_0 \gamma_{21} \phi_{0}^{-1} \rangle_1 +
(\partial\chi - eA) = m\cos{\beta} v ,
\end{equation}
\begin{equation}
\label{eq.86}
\langle \partial\phi_0 \gamma_{21} \phi_{0}^{-1} \rangle_3 =
\gamma_5 m\sin{\beta} v .
\end{equation}
Eq.(\ref{eq.85}) looks as a factorized Hamilton-Jacobi equation with
$\chi$ playing the role of the action function.
Indeed, consider the free case: $A = 0$. In this case we have
plane waves solutions for which $\partial\phi_0 = 0$ and $\cos{\beta}
= \pm 1$, and eq.(\ref{eq.85}) reduces to
$\partial\chi = \pm mv $,
which squares to $(\partial\chi)^2 = m^2 $,
which is the relativistic Hamilton-Jacobi equation. When we have
interactions, Hamilton-Jacobi equation is modified according to
$\partial\chi \mapsto \partial\chi - eA$ (which is just the term
in parenthesis on the LHS of eq.(\ref{eq.85})) and the factorized
Hamilton-Jacobi equation would assumes the form
$\partial\chi - eA = \pm mv$. Now, a comparison with eq.(\ref{eq.85})
shows that the mass term in that factorized Hamilton-Jacobi
equation is
\begin{equation}
\label{eq.87}
M = m\cos{\beta} ,
\end{equation}
which is just our variable mass given by eq.(\ref{eq.79}). On the
other hand, if mass appears as being variable, there must be
something that causes it to be so, and this seems to be the first
term on the LHS. The natural interpretation of this first term
is a quantum electromagnetic potential $A_{\scriptscriptstyle Q}$:
\begin{equation}
\label{eq.88}
e A_{\scriptscriptstyle Q} =
\langle \partial\phi_0 \gamma_{21} \phi_{0}^{-1} \rangle_1  .
\end{equation}
This interpretation looks very nice, but it remains the
problem of interpreting the other eq.(\ref{eq.86}). A possible
one consists in a quantum electromagnetic pseudo-potential
$\langle \partial\phi_0 \gamma_{21} \phi_{0}^{-1} \rangle_3$, but
it is not clear what underlies it in the context of magnetic
monopoles (where one can use two electromagnetic potentials -- a vector
and a pseudo-vector ones -- instead of the Dirac string \cite{Maia}).
But it is interesting to observe that in the Schr\"odinger limit
where $\phi_0$ is a real scalar the term $(\partial \phi_0)\phi_0^{-1}$,
when $\phi_0$ has no time dependence, is just the factorization
of the term $\nabla^2 \phi_0 + \mid \! \nabla \phi_0 /\phi_0 \! \mid^2
\phi_0$ (remember that $\partial^2 = \Box = \partial_0^2 - \nabla^2$),
which is introduced as nonlinearity in Schr\"odinger equation in
order to have an equation admitting quantum diffusion
currents \cite{Doebner}.

We can also speculate another way of interpreting eqs.(\ref{eq.85},
\ref{eq.86}). We saw in the preceeding section that one can
associate a spinor field $\psi$ to an electromagnetic field $F$
by means of $F = \psi \gamma_{21}\tilde{\psi}$. Our idea now
is to suppose $\phi_0$ associated to a ``quantum'' electromagnetic
field $F = \phi_0 \gamma_{21} \tilde{\phi}_0 $. We shall suppose
$F$ to satisfies the free Maxwell equations, and in this way $\phi_0$
satisfies eq.(\ref{eq.47}) with ${\cal J} = 0$. On the other hand,
in this section we imposed two constrains on the spinor field
$\psi$ since it has two additional degrees of freedom for
representing an electromagnetic field. Now, however, the
spinor $\phi_0$ has seven degrees of freedom (eq.(\ref{eq.82})), and
only one constrain has to be imposed. This constrain, however,
cannot be arbitrary; since $\phi\gamma_0 \tilde{\phi} =
\phi_0 \gamma_0 \tilde{\phi}_0 = \rho v = J$ and $\partial \cdot J = 0$,
this constrain must be
\begin{equation}
\label{eq.89}
\partial \cdot (\phi_0 \gamma_0 \tilde{\phi}_0) =
\partial \cdot (\rho v) = 0 .
\end{equation}
Now, eq.(\ref{eq.47}) with ${\cal J} = 0$ is (with
$\psi$ replaced by $\phi_0$, which we indicate
by the subscript):
\begin{equation}
\label{eq.90}
\partial\phi_0 \gamma_{21}\phi_{0}^{-1} =
\frac{{\rm e}^{\gamma_5 \beta}}{\rho}(j_0 + \gamma_5 g_0)  .
\end{equation}
Using eqs.(\ref{eq.53},\ref{eq.54}) we have eqs.(\ref{eq.57},\ref{eq.58})
for $j_0$ and $g_0$, which in the above equation gives
\begin{equation}
\label{eq.91}
\partial\phi_0 \gamma_{21}\phi_{0}^{-1} = {\rm e}^{\gamma_5 \beta}
(\lambda_0 + \gamma_5 \kappa_0)v = (\Lambda_0 + \gamma_5 K_0) v ,
\end{equation}
and then
\begin{equation}
\label{eq.92}
\langle \partial\phi_0 \gamma_{21}\phi_{0}^{-1} \rangle_1 =
\Lambda_0 v ,
\end{equation}
\begin{equation}
\label{eq.93}
\langle \partial\phi_0 \gamma_{21}\phi_{0}^{-1} \rangle_3 =
\gamma_5 K_0 v .
\end{equation}
Using these expressions in eqs.(\ref{eq.85},\ref{eq.86}) we
have ($\hbar = c = 1$):
\begin{equation}
\label{eq.94}
\Lambda_0 v + \partial\chi -eA = m\cos{\beta} v ,
\end{equation}
\begin{equation}
\label{eq.95}
K_0 v = m \sin{\beta} v .
\end{equation}
Finally, using the definitions of $\Lambda_0$ and $K_0$
we have
\begin{equation}
\label{eq.96}
\gamma^{\mu}(\Omega_{\mu}^{(0)}\cdot S) + \partial\chi -eA =
m\cos{\beta}v  ,
\end{equation}
\begin{equation}
\label{eq.97}
\gamma^{\mu}(\Omega_{\mu}^{(0)}\cdot(\gamma_5 S)) =
m\sin{\beta} v ,
\end{equation}
where by the superscript $(0)$ we indicate that
$\Omega_{\mu}^{(0)}$ is to be calculated
using $R_0$. Since $R = R_0 {\rm exp}(-\gamma_{21}\chi)$ we
have
\begin{equation}
\label{eq.98}
\gamma^{\mu}(\Omega_{\mu}\cdot S) = \gamma^{\mu}(\Omega_{\mu}^{(0)}\cdot S)
+ \partial\chi  ,
\end{equation}
\begin{equation}
\label{eq.99}
\gamma^{\mu}(\Omega_{\mu}\cdot(\gamma_5 S)) =
\gamma^{\mu}(\Omega_{\mu}^{(0)}\cdot(\gamma_5 S)) ,
\end{equation}
and we can write that
\begin{equation}
\label{eq.100}
m\cos{\beta} = \Omega \cdot S - e(A \cdot v) ,
\end{equation}
\begin{equation}
\label{eq.101}
m\sin{\beta} = \Omega \cdot (\gamma_5 S) ,
\end{equation}
which in the free case reduce to eqs.(\ref{eq.53},\ref{eq.54}) once we
identify $\Lambda$ with $m\cos{\beta} = M$ (as we already have done!)
and $K$ with $m\sin{\beta}$ (which is again the case due to
eq.(\ref{eq.69})!). This shows the consistency of this interpretation.

Finally, we observe that an interesting interpretation for $m \cos{\beta}$
and $m \sin{\beta}$ in eqs.(\ref{eq.85},\ref{eq.86}) and
eqs.(\ref{eq.100},\ref{eq.101}) would be that of longitudinal
and transversal masses, respectively. Further possible interpretations
along this line will be postponed for another occasion.

 \section{Conclusions}

In this paper we obtained a spinorial representation of Maxwell
equations and showed how it can be put in a form which is identical
to Dirac equation in the free case. We also discussed certain conditions
under which that equation can be really interpreted as Dirac equation.
These conditions are related to the interpretation of the mass
term, and we developed possible interpretations along some of
de Broglie ideas. The natural unit of mass, or the natural unit
of length, appears in our approach due to the theorem we proved which
gives eq.(\ref{eq.32}) relating electromagnetic and spinor fields.

A problem to be considered in another occasion is the case when we
have sources. In fact, the above relationship between Maxwell and
Dirac equations emerged from the free Maxwell equations. However,
we have a current $e\psi\gamma_0 \tilde{\psi}$ which is a source of
a self-field; that is, we have a feedback process. Studies in this
case can be found in \cite{Cam90}.

We also speculate on an alternative interpretation
by splitting the Dirac-Hestenes spinor in a phase given by the
action function and a spinor to which we associate a ``quantum''
electromagnetic field according to a theorem which we proved.
This scheme is consistent once that ``quantum'' electromagnetic
field satisfies the free Maxwell equations.

A possible and interesting generalization of our scheme consists
in introducing sources for the ``quantum'' electromagnetic field.
An interesting candidate for the current would be a self
electromagnetic potential. Indeed, interesting results can be
obtained (see, for example, \cite{RVR} and references therein) when
the electromagnetic potential
plays the role of current in Maxwell equations.

To end we must observe that our construction of the equivalence between
Mawell and Dirac equations depends on the existence of non trivial
solutions of the free Maxwell equations $\partial F=0$ such that
$F^2 \neq 0$. We proved in \cite{XXX} that non trivial solutions
in these conditions indeed exist. They correspond to subluminal and
superluminal solutions of Maxwell equations in vacuum!

\acknowledgements{We are grateful for discussions
to G. Cabrera, S. Ragusa, E. Recami, Q.~A.~G. Souza and W. Seixas.
One of the authors (J.V.) wishes to thank J. Keller
and A. Rodriguez for their very kind hospitality at
FESC - UNAM and UASLP, respectively.
This work has been partially supported by CNPq and FAEP-UNICAMP.}

\end{document}